\documentclass[aps,prc,nofootinbib,showpacs]{revtex4}

\usepackage{epsfig}
\usepackage{graphicx}

\begin{document}
\title{Possible methods for the determination of the $P$-parity of the $\Theta^+$-pentaquark in NN-collisions.}
\author{Michail P. Rekalo }
\affiliation{National Science Centre - Kharkov Institute of 
Physics and Technology,\\ Akademicheskaya 1, 61108 Kharkov,
Ukraine}
\author{Egle Tomasi-Gustafsson}
\affiliation{\it DAPNIA/SPhN, CEA/Saclay, 91191 Gif-sur-Yvette Cedex, 
France}
\date{\today}


\begin{abstract}
We present two possibilities to determine the P-parity of the pentaquark $\Theta^+$, in a model independent way, via the measurement of polarization observables in $p+p\to \Theta^+ +\Sigma^+$, or $n+p\to \Theta^+ +\Lambda^0$,
in the near threshold region. Besides the measurement of the spin correlation coefficient, $A_{xx}=A_{yy}$, (in collisions of transversally polarized nucleons), the coefficient $D_{xx}$ of polarization transfer from the initial proton to the final $ \Sigma^+(\Lambda^0)$ hyperon is also unambiguously related to the $\Theta^+$ parity.
\end{abstract}
\maketitle
In 1999, N. K. Pak and M. Rekalo proposed \cite{Pa99} two new methods for the determination of the P-parity of the $K$-meson, through the measurement of different polarization observables in $K$-meson production in proton proton collisions near threshold, $p+p\to K^++Y^0+p$ ($Y^0=\Lambda$ or $\Sigma$-hyperon). One method is based on the measurement of the sign of the spin correlation coefficient $A_{yy}$, for collisions of transversally polarized protons. The second one is based on the measurement of the polarization transfer coefficient from the initial proton to the produced hyperon, $D_{nn}$. Both methods apply in threshold conditions, where all final particles are in S-state. Note, in this respect, that the DISTO collaboration  showed the feasibility of the second method, by measuring the $D_{nn}$ coefficient at proton momentum of 3.67 GeV/c, which showed that {\it "$D_{nn}$ is large and negative ($\simeq -0.4$) over most of the kinematic region"} \cite{DISTO}. It was mentioned in \cite{Pa99} that a nonzero value of $D_{nn}$, in the threshold region, can be considered as the experimental confirmation of the pseudoscalar nature of the $K^+$ meson.

It is straightforward to adapt these methods to the determination of the P-parity of the $\Theta^+$-hyperon, which is presently object of an intensive theoretical discussion. 

The simplest reactions of $NN$-collisions, which can be considered for this aim are the following:
\begin{eqnarray}
p +n &\to &\Lambda^0 +\Theta^+,\label{eq:reac1} \\
p+ p &\to &\Sigma ^+ +\Theta^+, \label{eq:reac2}  \\
p+ p &\to &\pi^+ +\Lambda^0 +\Theta^+,\label{eq:reac3}
\end{eqnarray}
in threshold conditions (S-wave production). It is important to note that the $\Lambda^0$ or the $\Sigma ^+$ hyperons, produced in these reactions, are self analyzing particles, therefore the polarization transfer method - with measurement of the polarization transfer coefficient $D_{nn}$ seems more preferable, from the experimental point of view.
 
The following analysis of reactions (1-3) is based on general symmetry properties of the strong interaction, such as the P-invariance, the conservation of the total angular momentum, the Pauli principle, for the $pp$-system, and the generalized Pauli principle for the $np$ system, which holds at the level of the isotopic invariance of the strong interaction. 

These symmetry properties, being applied to S-wave production in the processes (1-3) allow to establish the spin structure of the corresponding matrix elements, for both cases of the considered P-parity. 

We consider here, for simplicity,  the case of spin 1/2 $\Theta^+$ hyperon, but this formalism can be extended to any $\Theta^+$ spin.

Firstly, let us establish the general spin structure of double polarization observables for the processes (1-3) at threshold.

The dependence of the cross section, total or differential, on the polarizations $\vec P_1$ and $\vec P_2$ of the colliding nucleons can be written as:
\begin{equation}
\sigma(\vec P_1,\vec P_2)=\sigma_0(1+{\cal A}_1\vec P_1\cdot\vec P_2+{\cal A}_2\hat{\vec k}\vec P_1\cdot\hat{\vec k}\vec P_2),
\label{eq:eqp}
\end{equation}
where $\sigma_0$ is the cross section for the collision of unpolarized nucleons, $\hat{\vec k}$ is the unit vector along the three momentum of the colliding nucleons, in the reaction CM system. The real coefficients ${\cal A}_1$ and 
 ${\cal A}_2$, which are different for different reactions, depend on the parity of the $\Theta^+$ hyperon. Taking the $z$-axis along   $\hat{\vec k}$, one can find the following expression for the spin correlation coefficients ${\cal A}_{ab}$, in terms of ${\cal A}_1$ and ${\cal A}_2$:\
\begin{equation} 
{\cal A}_{xx}={\cal A}_{yy}={\cal A}_1,~{\cal A}_{zz}={\cal A}_1+{\cal A}_2.
\label{eq:eq5}
\end{equation}
The dependence of the polarization $\vec P_Y$ of the produced hyperon, $\Lambda$ or $\Sigma^+$, on the polarization $\vec P$ of the initial nucleon (beam or target) can be written as:
\begin{equation} 
\vec P_Y=p_1\vec P + p_2\hat{\vec k}~\hat{\vec k}\cdot\vec P,
\label{eq:eq6}
\end{equation}
where $ p_{1,2}$ are real coefficients, which depend on the $\Theta^+$ parity, so that for the non-zero polarization transfer coefficients, $D_{ab}$ one can write:
\begin{equation} 
{\cal D}_{xx}={\cal D}_{yy}=p_1,~{\cal D}_{zz}=p_1+p_2.
\label{eq:eq7}
\end{equation}

Let us calculate these coefficients for the reactions (1-3), it terms of S-wave partial amplitudes, considering both values of the $\Theta^+$ parity.

\noindent\underline{\bf $n+p\to \Lambda+\Theta^+$}. This reaction has the lowest threshold, and seems very interesting for the measurement of the ${\cal D}_{nn}$ coefficient, due to the large asymmetry and branching ratio of the decay $\Lambda\to p+\pi^-$ ($\alpha=0.642\pm 0.013$ and Br=$(63.9\pm 0.5)$\% \cite{PDG}).

The spin structure of  the threshold matrix element depends on the discussed P-parity (assuming that the isotopic spin of $\Theta^+$ is zero):
\begin{equation}
{\cal M}^{(-)}_{\Lambda}=f^{(-)}(\Lambda) {\cal I}\bigotimes \vec\sigma\cdot\hat{\vec k},~\mbox{~if~}P(\Theta^+)=-1,
\label{eq:eq8}
\end{equation}
\begin{equation}
{\cal M}_{\Lambda}^{(+)}=f_1^{(+)}(\Lambda) \vec\sigma\cdot\hat{\vec k}\bigotimes \vec\sigma\cdot\hat{\vec k}+f_2^{(+)}(\Lambda)(\sigma_m-\hat k_m 
\sigma\cdot\hat{\vec k})\bigotimes \sigma_m,~\mbox{~if~}P(\Theta^+)=+1
\label{eq:eq9}
\end{equation}
where the upper index $(\pm)$ for the partial amplitudes $f(\Lambda)$ corresponds to $P(\Theta^+)=\pm 1$. Henceforward we use the following abbreviation
\begin{equation}
A\bigotimes B =(\tilde \chi_2 \sigma_yA\chi_1)( \chi_4^{\dagger} B\sigma_y\tilde \chi_3^{\dagger}),
\label{eq:eq10}
\end{equation}
where $\chi_1$ and $\chi_2$ ($\chi_3$ and $\chi_4)$ are the two-component spinors of the initial (final) baryons.

Using Eqs. (\ref{eq:eq8}) and (\ref{eq:eq9}) one can find the following formulas for double spin polarization observables:
\begin{equation}
{\cal A}_{xx}^{(-)}(\Lambda)={\cal A}_{yy}^{(-)}(\Lambda)=
{\cal A}_{zz}^{(-)}(\Lambda)=-1,~
{\cal D}_{ab}^{(-)}=0,~\mbox{~if~}P(\Theta^+)=-1, 
\label{eq:eq12}
\end{equation}
and
\begin{equation}
D_{\Lambda}^{(+)}{\cal A}_{xx}^{(+)}(\Lambda)=
D_{\Lambda}^{(+)}{\cal A}_{yy}^{(+)}(\Lambda)=
|f_{1\Lambda}^{(+)}|^2,~
D_{\Lambda}^{(+)}{\cal A}_{zz}^{(+)}(\Lambda)=
2\left (-|f_1^{(+)}(\Lambda)|^2+|f_2^{(+)}(\Lambda)|^2\right)
\label{eq:eq13}
\end{equation}
\begin{equation}
D_{\Lambda}^{(+)}{\cal D}_{xx}^{(+)}(\Lambda)=
D_{\Lambda}^{(+)}{\cal D}_{yy}^{(+)}(\Lambda)=
2 Re f_{1\Lambda}^{(+)}f_{2\Lambda}^{(+)*},~
D_{\Lambda}^{(+)}{\cal D}_{zz}^{(+)}(\Lambda)=
2|f_2^{(+)}(\Lambda)|^2,
\label{eq:eq14}
\end{equation}
with 
$$
~D_{\Lambda}^{(+)}=|f_1^{(+)}(\Lambda)|^2+2|f_2^{(+)}(\Lambda)|^2,
~\mbox{~if~}P(\Theta^+)=+1,
$$
So, comparing the two possibilities for the P-parity, one can predict, in model independent way:
$$
{\cal A}_{xx}^{(-)}(\Lambda)={\cal A}_{yy}^{(-)}(\Lambda)=-1,~
{\cal D}_{yy}^{(-)}(\Lambda)=0,\mbox{~if~} P(\Theta^+)=-1,
$$
\begin{equation}
{\cal A}_{yy}^{(+)}(\Lambda)\ge 0, ~{\cal D}_{yy}^{(-)}(\Lambda)\ne 0,
\mbox{~if~} P(\Theta^+)=+1
\label{eq:eq15}
\end{equation}
with evident sensitivity of these observables to the parity of the $\Theta^+$ hyperon.


\noindent\underline{\bf $p+p\to \Sigma^++\Theta^+$}\footnote{The collisions of polarized protons in this reaction have been considered in \protect\cite{Th03}.}.
The spin structure of the threshold matrix element is different from $\Lambda$ production ( due to the difference in the value of the total isotopic spin for the colliding nucleons) and depends on $P(\Theta^+)$:
\begin{equation}
{\cal M}_{\Sigma}^{(+)}=f^{(+)}(\Sigma) {\cal I}\bigotimes {\cal I},~\mbox{~if~}P(\Theta^+)=+1
\label{eq:eq17}
\end{equation}
\begin{equation}
{\cal M}_{\Sigma}^{(-)}=f_1^{(-)}(\Sigma) \vec\sigma\hat{\vec k}\bigotimes {\cal I} 
+if_2^{(-)}(\Sigma)(\vec\sigma\times\hat{\vec k})_m) \bigotimes \sigma_m,~\mbox{~if~}P(\Theta^+)=-1
\label{eq:eq16}
\end{equation}
where $f^{(\pm)}(\Sigma)$ are the corresponding partial amplitudes for $P(\Theta^+)=\pm 1$, in case of 
$\Sigma$ production.

Using these matrix elements, one can find the following form for the corresponding double polarization observables:
$$D_{\Sigma }^{(-)}{\cal A}_{xx}^{(-)}(\Sigma )=
D_{\Sigma }^{(-)}{\cal A}_{yy}^{(-)}(\Sigma )=|f_1^{(-)}(\Sigma)|^2,$$
$$D_{\Sigma }^{(-)}{\cal A}_{zz}^{(-)}
(\Sigma )=-|f_1^{(-)}(\Sigma)|^2+2|f_2^{(-)}(\Sigma)|^2,$$
\begin{equation}
D_{\Sigma }^{(-)}{\cal D}_{xx}^{(-)}(\Sigma )=
D_{\Sigma }^{(-)}{\cal D}_{yy}^{(-)}(\Sigma )=2Re f_1^{(-)}(\Sigma)f_2^{(-)*}(\Sigma),
\label{eq:eq18}
\end{equation}
$$D_{\Sigma }^{(-)}{\cal D}_{zz}^{(-)}(\Sigma )=2|f_1^{(-)}(\Sigma)|^2,$$
with $D_{\Sigma }^{(-)}=|f_1^{(-)}(\Sigma)|^2+2|f_1^{(-)}(\Sigma)|^2$, in case of negative parity of the $\Theta^+$, and 
\begin{equation}
{\cal A}_{xx}^{(+)}(\Sigma )={\cal A}_{yy}^{(+)}(\Sigma )={\cal A}_{zz}^{(+)}(\Sigma )=-1,~
{\cal D}_{ab}^{(+)}(\Sigma )=0
\label{eq:eq19a}
\end{equation}
in case of positive parity.

Therefore, the measurement of the quantities ${\cal A}_{yy}(\Sigma )$ and ${\cal D}_{yy}(\Sigma )$ allows to determine the P-parity:
$${\cal A}_{yy}^{(-)}(\Sigma )\ge 0,~{\cal D}_{yy}^{(-)}(\Sigma )\ne 0~\mbox{~if~}P(\Theta^+)=-1
$$
\begin{equation}
{\cal A}_{yy}^{(+)}(\Sigma )=-1,~{\cal D}_{yy}^{(+)}(\Sigma )=0 ~\mbox{~if~}P(\Theta^+)=+1.
\label{eq:eq19}
\end{equation}

Comparing Eqs. (\ref{eq:eq15}) and (\ref{eq:eq19}), one can see that the reactions of $\Theta^+$ production in $NN$-collisions, $ p+p \to \Sigma ^+ +\Theta^+$ and $ n+p \to \Lambda^0 +\Theta^+$, which look similar at first sight, show a very different dependence of the 
${\cal A}_{yy}$ and ${\cal D}_{yy}$  observables on the P-parity of the $\Theta^+$ hyperon. For example, the signs of ${\cal A}_{yy}(\Sigma )$ and ${\cal A}_{yy}(\Lambda)$ asymmetries are different, independently on $P(\Theta^+)$. A large difference is also present in the ${\cal D}_{yy}(\Sigma )$ and  ${\cal D}_{yy}(\Lambda)$ observables.
\noindent\underline{\bf $p+p+\to \pi^+ +\Lambda+\Theta^+$.} The spin structure of the corresponding matrix elements can be written as follows:
$$
{\cal M}^{(-)}=f^{(-)} {\cal I}\bigotimes{\cal I},~\mbox{~if~}P(\Theta^+)=-1
$$

\begin{equation}
{\cal M}^{(+)}=f_1^{(+)} \vec\sigma\cdot\hat{\vec k}\bigotimes 
{\cal I}+if_2^{(+)}(\vec\sigma\times\hat{\vec k})_m \bigotimes \sigma_m,~\mbox{~if~}P(\Theta^+)=+1,
\label{eq:eq20}
\end{equation}
i.e. similar to the reaction $ p+p \to \Sigma ^+ +\Theta^+$, but with opposite parity. So, for $p+p\to\pi^+ +\Lambda+\Theta^+$, the necessary polarization observables can be described by Eqs.(\ref{eq:eq18}) and (\ref{eq:eq19a}), taking care to interchange the P-parities.

This analysis shows that all the reactions (1-3) are well adapted to the determination of $P(\Theta^+)$. In all these reactions two polarization observables, namely ${\cal A}_{yy}$ and ${\cal D}_{yy}$  are sensitive to this parity. The signature of the parity of $(\Theta^+)$ is the sign of the asymmetry  ${\cal A}_{yy}$, or a a value  of the transfer polarization tensor, different from zero. These statements are model independent.

\begin{figure}
\mbox{\epsfxsize=15.cm\leavevmode \epsffile{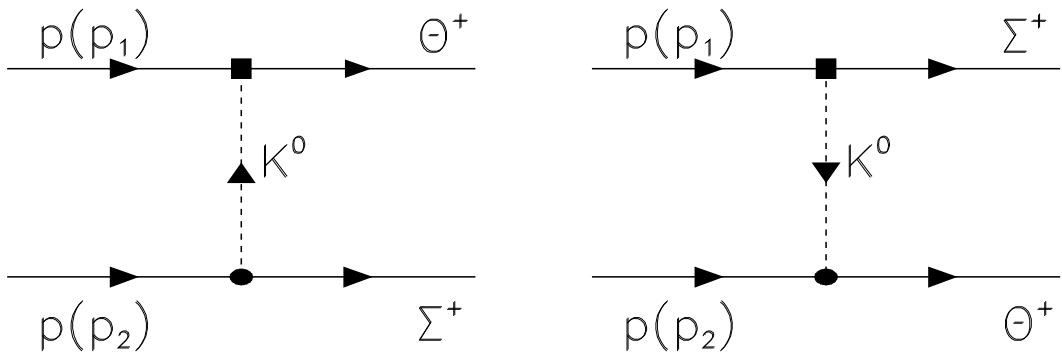}}
\caption{ $K$-exchange for the reaction $ p+p \to \Sigma ^+ +\Theta^+$.
}
\label{fig:fig1}
\end{figure}
Let us briefly discuss the expected values of the polarization observables, in the reactions (1-3), which depend on two amplitudes, $f_1$ and $f_2$. For this aim, we will take, as an example, a model of $K$-meson exchange, which has been applied to the $\Lambda$ and $\Sigma^0$ production in $pp$-collisions \cite{Mo02}, and reproduces quite well the sign and the absolute value of  $D_{nn}$ in $\vec p+p\to \vec\Lambda+K^++p$. Considering the contribution of both diagrams in Fig. 1, one can find:
\begin{equation}
f_1^{(-)}(\Sigma)=-f_2^{(-)}(\Sigma)
\label{eq:eq21}
\end{equation}
This relation does not depend on many details of the reaction mechanism, such as the values of the two coupling constants, $g_{p\Theta K}$ and $g_{p\Sigma K}$, on the width of $\Theta^+ $ and on the form of the phenomenological form factors, which has to be taken into account in such considerations.

The relation (\ref{eq:eq21}) allows to predict:
\begin{equation}
{\cal A}_{yy}^{(-)}(\Sigma )=+1/3,~{\cal D}_{yy}^{(-)}(\Sigma )=-2/3.
\label{eq:eq22}
\end{equation}
Let us note that $|{\cal D}_{yy}^{(-)}(\Sigma )|$  is different from zero and large. This will make easier to discriminate the value of the P-parity.

The same relation holds also for the amplitudes $f_{1,2}^{(+)}(\Lambda)$ for the process 
$n+p\to\Lambda+\Theta^+$ (for $K$-exchange), with corresponding predictions for polarization effects.

Note, however, that final state $\Sigma\Theta$- interaction, which is different, generally, in singlet and triplet states, can affect the relation (\ref{eq:eq21}). There are arguments which show that these effects can not be large \cite{Th03}. In any case we considered here $K$-exchange only for illustrative purposes, for a quick estimation of polarization phenomena, without any claim that this is a realistic model for these reactions \cite{Na04}: the main result of this paper does not depend on model considerations. 

The experimental study of all three reactions (1-3) will give a non ambiguous signature of the $\Theta^+$ parity.

 
{}

\end{document}